\documentclass[english]{lni} %[anonymous] for review
\usepackage{booktabs}

\usepackage{tabularx}
\usepackage{multirow}

\usepackage{caption}    %für die Mehrfach-Graphiken
\DeclareCaptionFormat{bold}{\textbf{#1#2}#3}
\begin{document}
%%% Das optionale Argument (sofern angegeben) wird für die Kopfzeile verwendet.
\title{Evaluating Software Supply Chain Security in Research Software}
\subtitle{An Empirical Assessment of 3,248 Repositories Using OpenSSF Scorecard } 
\author[1]{Richard Hegewald}{richard.hegewald@uni-potsdam.de}{0009-0006-3953-6445}
\author[1]{Rebecca Beyer}{rebecca.beyer@uni-potsdam.de}{0009-0005-3204-4459}
\affil[1]{University of Potsdam\\Institute of Computer Science\\An der Bahn 2\\14476 Potsdam\\Germany}
\maketitle

(Author's version accepted for GI SKILL 2025)

\begin{abstract}
The security of research software is essential for ensuring the integrity and reproducibility of scientific results. However, research software security is still largely unexplored. Due to its dependence on open source components and distributed development practices, research software is particularly vulnerable to supply chain attacks. This study analyses 3,248 high-quality, largely peer-reviewed research software repositories using the OpenSSF Scorecard. We find a generally weak security posture with an average score of 3.5/10. Important practices, such as signed releases and branch protection, are rarely implemented. Finally, we present actionable, low-effort recommendations that can help research teams improve software security and mitigate potential threats to scientific integrity.

\end{abstract}


\section{Introduction}

    The increasing prevalence of software supply chain attacks presents a major challenge to cybersecurity across all sectors. In the open-source ecosystem alone, over 512,000 malicious packages were identified in the past year, marking a 156\% year-over-year increase \cite{Sonatype2024}. \\
    Research Software (RS) constitutes a common type of open-source software, encompassing ''source code files, algorithms, scripts, computational workflows and executables that were created during the research process or for a research purpose'' \cite{Gruenpeter2021}.
    Since RS is an integral part of scientific research, ensuring its security is critical to maintaining confidence in scientific results. Compromised research software can undermine both the integrity and reproducibility of results, create risks for sensitive data and violate the FAIR principles (Findable, Accessible, Interoperable and Reusable) by limiting reusability \cite{Barker2022}. 
    From a security perspective, RS is particularly vulnerable due to the high number of dependencies of packages, which creates a broad attack surface and provides numerous vectors for compromise \cite{Prana2021}. Despite this, the security status of research software repositories has not yet been systematically assessed, leaving a gap in our understanding of the risks to the scientific community.\\
    A compromise in the software supply chain of research software is not only a theoretical risk. Researchers who installed PyTorch\footnote{specifically: pytorch-nightly installed via pip on Linux}, a foundational Python package for AI research, between the 25th and 30th December 2022, not only downloaded the library itself, but also the malicious Triton binary. Triton scanned and exfiltrated sensitive information from the host machine. This attack did not target the source code of the PyTorch directly, but merely replaced a dependency with a malicious package \cite{Pytorch2022}.
    While research software generally has the same risk of falling prey to unspecific attacks as other open-source packages, this case presents a targeted attack on a major research software package. This incident is not isolated, as evidenced by \citet{Conda2025, Sharma2024}, and as such raises questions about the overall resilience of research software against future attacks. \\
    Previous large-scale analyses have focused on entire package ecosystems, but these approaches often overlook the particular constraints faced by RS engineers, such as the lack of formal software engineering training and limited resources for maintenance \cite{Johanson2018, Anzt2021}. Therefore, the results of such studies can be difficult to interpret or apply to the context of academic research, where operational realities differ from those of commercial software development.
    In this study, we conduct a domain-focused analysis of the security posture of largely peer-reviewed research software. Our analysis is based on the OpenSSF Scorecard tool and targets publicly available GitHub repositories to answer the following research questions:
    
    \textit{(RQ1) What is the current state of supply chain security in research software?} \\
    \textit{(RQ2) What security best practices are commonly adopted in research software, and which are often overlooked or underutilised?}\\
    \textit{(RQ3) What actionable recommendations can be made to research software engineers to improve the supply chain security of their software?}

\section{Background and Related Work}
% Which threats have supply chain software ?
% What are possible methods to asses security status of software
% Was ist research Software ENginering, was typisch dafür?

    % Scorecard
        The \textbf{OpenSSF Scorecard} is a tool for automatically assessing the security of git repositories. It was launched in November 2020 and is maintained and organised by the OpenSSF Best Practices Working Group. Scorecard is organised into a series of checks that can be run independently on a software repository. At the time of writing, Scorecard implements 20 checks, 18 of them being run by default, while the remaining 2 are considered experimental. Checks are categorised into 4 risk levels (Low, Medium, High, and Critical) and scored using a scale from 0 to 10, with 0 being the lowest and 10 being the highest security score. An aggregate score is calculated by adding up the individual check scores, weighted by risk level \cite{OSSF_Scorecard}.
        Recently \citet{Zahan2023a} presented a large scale analysis of the PyPI and npm package ecosystems using the OpenSSF Scorecard, pointing out relevant gaps in the adoption of security best practices. While ecosystem wide analysis can be helpful in assessing the general state of security in these packages, this approach separates software projects mainly by the employed programming language and fails to take into account different types, scopes and resources of software projects. Scorecard has also been used to assess the effect of various security best practices on vulnerability count in \citet{Zahan2023b}. However, the predictive power of this approach was limited, as noted by the authors. Furthermore, \citet{Younis2023} use Scorecard to provide a domain-focused analysis on the supply chain security of industrial control system protocols. \\
    %SLSA
         \textbf{Supply-chain Levels for Software Artifacts} (SLSA) presents a lightweight, end-to-end security framework initiated in 2021 by the Open Source Security Foundation.
         \cite{slsa}. The framework encompasses an accessible threat model and a set of security controls organised into tracks that organisations and developers can adopt to increase the supply chain security of their software \cite{slsa}. Since both SLSA and the Scorecard are organised by the OpenSSF and tailored to open-source projects, the integration of these two components provides a more suitable representation of the addressed threats, in comparison to other frameworks like OWASP Software Component Verification Standard as used in \citet{Zahan2023a}. In Table \ref{tab:slsa_mapping_threads}, we provide a mapping that presents which threat of the SLSA framework is addressed by each Scorecard check, showing that the OpenSSF Scorecard is an adequate tool for assessing software supply chain security.  \\

\begin{table}[ht]
    \footnotesize
    \centering
    \captionsetup{font=footnotesize}
    %\rowcolors{2}{gray!15}{white}
    \begin{tabular}{lp{8cm}}
        \toprule
        %\rowcolor{white}
        \textbf{SLSA Threat} & \textbf{Associated Checks} \\
        \midrule
        (A) Producer & Security-Policy, CI-Test, CII-Best-Practices, Contributors, License, Code-Review\\
        (B) Source Control & Binary-Artifacts, Branch-Protection, Code-Review, Contributors, Token-Permission \\
        (C) Source Platform & Token Permission \\
        (D) External Build Parameters & Dangerous-Workflow, Pinned-Dependencies \\
        (E) Build Process & Dangerous-Workflow, Packaging \\
        (F) Artifact Publication & Signed-Releases, Packaging \\
        (G) Distribution & Signed-Releases \\
        (H) Package Selection & Dependency-Update-Tool, Pinned-Dependencies \\
        (I) Usage & Vulnerabilities, Fuzzing, SAST, Maintained, CI-Test \\
        \bottomrule
    \end{tabular}
    \caption{\textbf{Assignment of the OpenSSF Scorecard Checks to the SLSA Threats according to SLSA v1.1}}
   \label{tab:slsa_mapping_threads}
\end{table}

\section{Methodology}
    
    \subsection{Data}
    One of the difficulties in assessing the security posture of research software projects is the variability in scope of each project. For our analysis, we aimed to include only projects that justify a certain commitment to software security, even to researchers unaware of supply chain risks. 
    % RSD or JOSS
    We decided on projects published in the Research Software Directory (RSD) of the Netherlands eScience Center \footnote{https://research-software-directory.org/ (accessed 20/07/2025)} and the Journal of Open Source Software (JOSS) \footnote{https://joss.theoj.org/ (accessed 20/07/2025)} to ensure a broad and representative sample of high quality research software from all disciplines to minimise selection bias.
    % RSD
    The Research Software Directory is a platform that allows researchers to register their software to showcase the impact of their software. The project provides a well documented API and is used by over 500 organisations.
    % JOSS
    The second and dominant source we included is software published in The Journal of Open Source Software. JOSS is an established software journal with a formal peer review process for every publication, including an extensive code review. Furthermore, according to the review criteria, JOSS contributions must represent a „substantial scholarly effort“, meaning „not less than three months of work for an individual“ \footnote{JOSS Review Criteria: https://joss.readthedocs.io/en/latest/review\_criteria.html (accessed 20/07/2025)}.

    \subsection{Data Collection \& Inclusion Criteria}
\begin{figure}
        \centering
        \captionsetup{font=footnotesize}
        \includegraphics[width=0.5\linewidth]{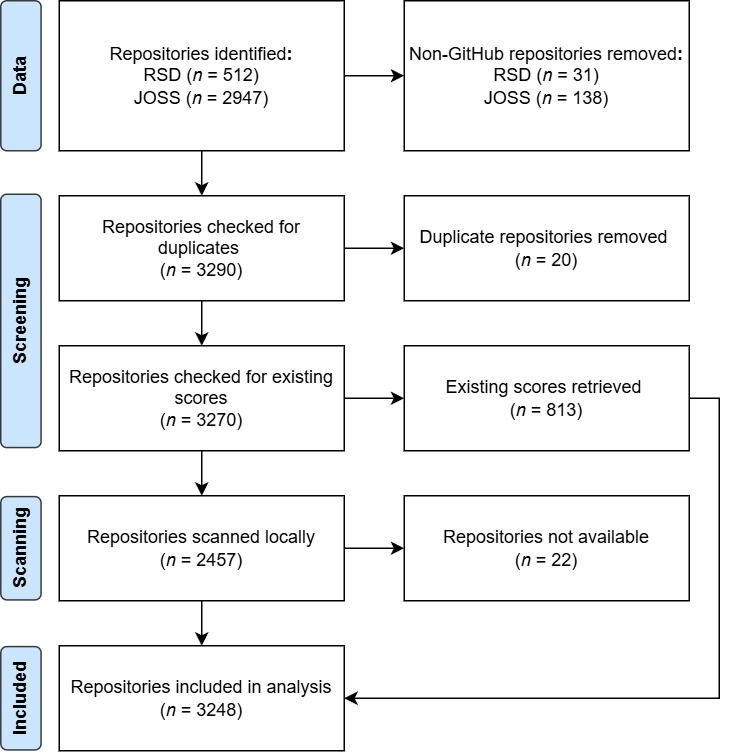}
        \caption{\textbf{Data Collection Workflow} based on the PRISMA 2020 guidelines \cite{Prisma2021}.}
        \label{fig:workflow}
    \end{figure}

    The data collection workflow is described in Figure \ref{fig:workflow}. To run a scan, OpenSSF Scorecard requires the URL of a software repository. For projects registered on the RSD, the corresponding repository URLs could be retrieved from REST API data \footnote{RSD REST API: https://research-software-directory.org/swagger/ (accessed 20/07/2025)}. Since there is no comprehensive list of repositories of JOSS contributions, we downloaded publicly available peer review discussions organised in the issues section of a central GitHub repository\footnote{JOSS Reviews: https://github.com/openjournals/joss-reviews/issues (accessed 20/07/2025)} using the GitHub API\footnote{Version 2022-11-28}. By filtering for issues with the \textit{accepted} tag, we were able to scrape the repository URL of each published, peer-reviewed contribution from the initial issue description.
    While the majority of collected repositories were hosted on GitHub, some were hosted on other source code management (SCM) platforms like GitLab. Although most of the Scorecard checks work on GitLab repositories as well, others are exclusive to GitHub. We therefore excluded any projects not hosted on GitHub from our analysis.
    After removing any duplicate repositories, we then queried the resulting list against the approximately 1.3M scan results which are reported weekly by the OpenSSF \cite{BigQueryScorecard2025}. If a score could be retrieved, it was downloaded and the repository removed from the list. The security scores of the remaining repositories were then calculated using a local instance of the OpenSSF Scorecard scanner with default configuration (for 18 security checks) during April 2025.

    \subsection{Preprocessing \& Analysis}
    
    The Scorecard results were preprocessed and analysed with Python using the NumPy \cite{NumPy} and Pandas \cite{Pandas} packages. 
    Data preprocessing included filtering out inconclusive results (scores of -1) for each check, ensuring that only interpretable and comparable scores were included in subsequent analyses.
    Descriptive statistics, including mean, standard deviation, and skewness, were determined for both aggregate and individual high risk security check scores. Individual checks were scored either on an ordinal scale ranging from 0 to 10 or on a binary scale (0 or 10), and all scores were included in the statistical analyses.
    Score distributions were visualised using Seaborn \cite{Seaborn}, SciPy \cite{SciPy} and Matplotlib \cite{Matplotlib} packages. The workflow itself was implemented in Snakemake \cite{Molder2021}.

    \subsection{Recommendations}
    Recommendations for improving supply chain security were derived by independently assessing each security check using a risk-adoptability matrix by the authors. The risk dimension for each security check was adopted directly from the risk classification provided by the OpenSSF Scorecard, which assigns risk levels to the individual checks. The adoptability assessment considered, for each practice, the expected ease of implementation for research software engineers, accounting for maintenance effort, implementation complexity, and required time. The individual assessments were subsequently consolidated into a single consensus matrix after jointly discussing individual assessments. Finally, three recommendations were prioritised with the aim of maximising security improvements by selecting those practices from the matrix that combine high security impact with high adoptability.

\section{Results}
     %Introduction Sentence for Results?

    \begin{figure}
                \centering
                \captionsetup{font=footnotesize}
                \includegraphics[width=0.8\columnwidth]{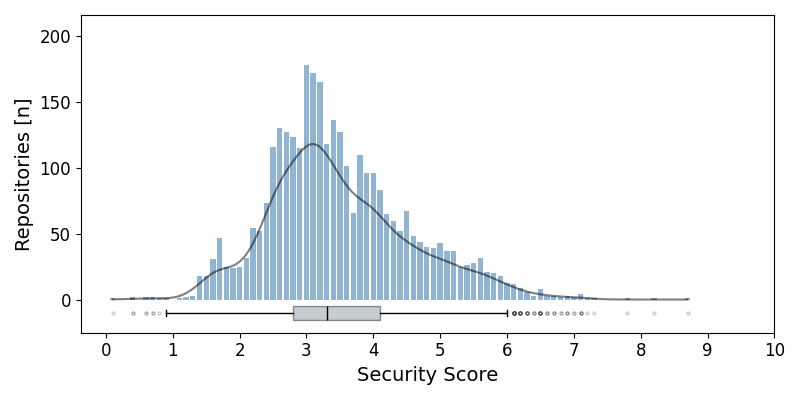}
                \caption{ \textbf{Aggregate Security Scores.} Distribution of aggregated Scorecard scores of 3248 research software repositories. Whiskers represent $\pm1.5$ IQR.}
                \label{fig:general_scores}
    \end{figure}

\begin{figure}[t]
    \centering
    \captionsetup{font=footnotesize}
    \includegraphics[width=1\linewidth]{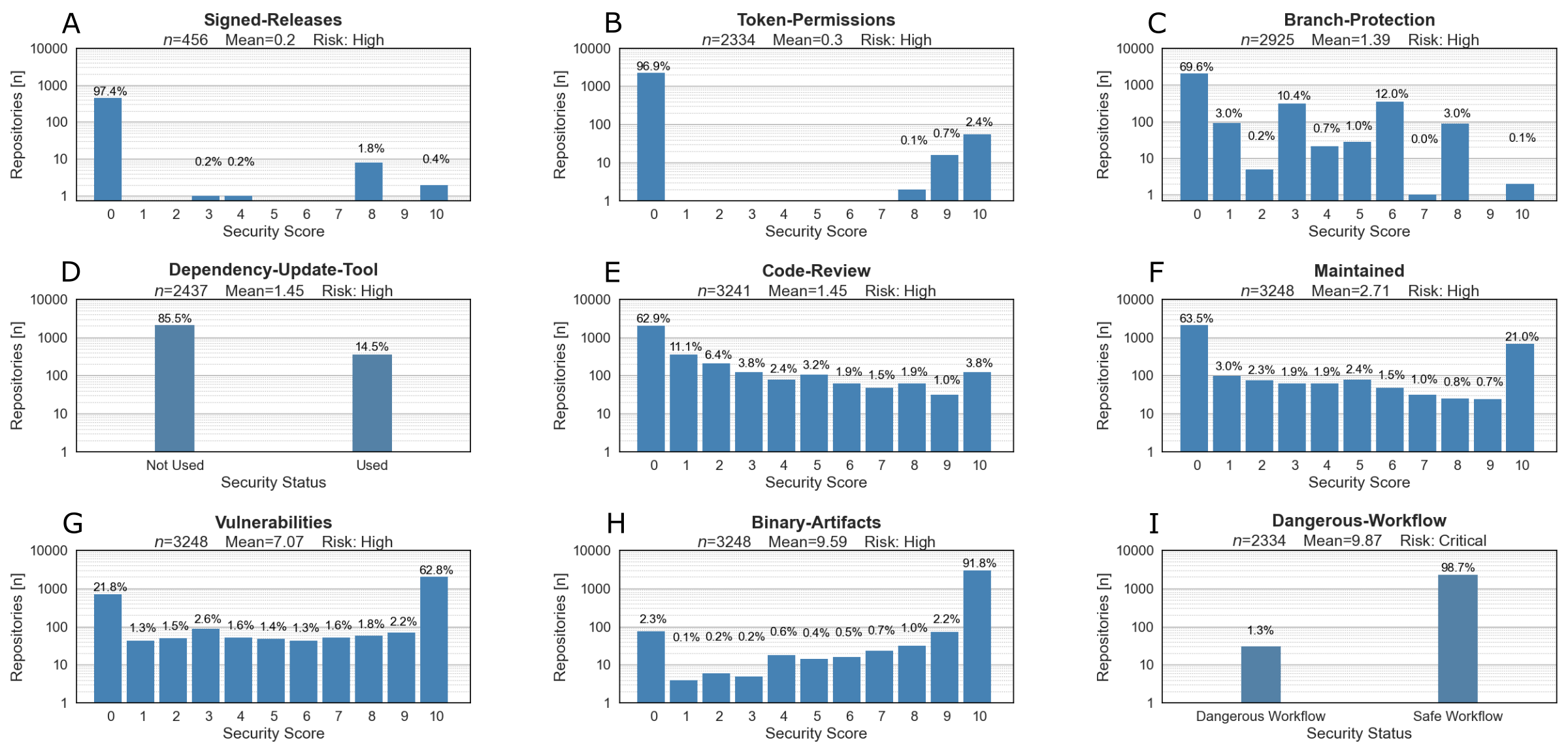}
    \caption{\textbf{Security State of individual high risk checks across repositories}. (A-I) Distributions of individual OpenSSF Scorecard check scores for high-impact security practices across the analysed repositories, sorted by ascending mean of scores of individual check. For each check sample size n and risk classification is given.}
    \label{fig:single_checks}
    \end{figure}

    %  Analyse General Score
    Figure \ref{fig:general_scores} presents the high-level results of our analysis of 3248 research software repositories as a distribution of aggregated Scorecard scores. The analysis yielded a mean aggregate score of $\mu=3.50$ with a standard deviation of $\sigma=1.06$. This indicates a substantial potential for improvement across the majority of scanned software repositories. Furthermore, 68\% of repositories score between $[2.43; 4.58]$ while 95\% fall within the range of $[1.35; 5.65]$. The resulting distribution has a significant positive skew $\gamma=0.62$ ($p < 3.7\cdot10^{-40}$). 
    %Individual Checks
    When examining the individual checks, substantial heterogeneity between the individual scores was observed. Of the 18 checks examined, 4 had a median score of 10, while 13 checks had a median score of 0. Overall, 13 out of 18  checks had a mean score below 4 (out of 10), whereas 3 checks had mean score above 7 (Tab. \ref{tab:security-checks}). Among the high-risk checks, 6 out of 9 had mean score below 3 (Fig. \ref{fig:single_checks}). \\
    % Signed Release
    \textbf{Signed-Releases} determines if a project cryptographically signs release artifacts to guarantee that the release artifacts have not been tampered with (Fig. \ref{fig:single_checks}A). Only for 14.04\% of those repositories a release could be found, with 97.4\% of those scoring 0, indicating that the majority of projects do not implement release signing. \\ 
    % Token-Permission
    \textbf{Token-Permission} evaluates whether a project's automated workflows tokens are set to read-only by default, limiting potential damage from token compromise (ex. prevent unauthorised repository modification). For 71.86\% of repositories a result could be found, for 96.9\% of those the set tokens were not set with the minimum required permissions (Fig. \ref{fig:single_checks}B).  \\
    % Branch-Protection
    \textbf{Branch-Protection} checks whether a project's default and release branches are protected with rules that enforce code review workflows, and prevent force pushes. 69.6\% of repositories scored 0, while only 0.1\% received the maximum score of 10; overall scores are concentrated on the lowest scores, with a mean of 1.39 (Fig. \ref{fig:single_checks}C). 27\% of the repositories prevented force push and branch deletion. Only 15.1\% require at least 1 reviewer approval before merging, and had admin exceptions: they required a pull-request prior to make any code changes, require branch to be up to date before merging and needed approval for most recent push. \\
    % Dependency-Update Tool
    \textbf{Dependency-Update-Tool} evaluates whether a project uses automated tools to update dependencies. For 75.03\% of the repositories a score of either 0 or 10 could be calculated. 85.5\% of scored repositories did not use an automatic dependency update tool and therefore received 0 points, while 14.5\% had such a tool enabled and received 10 points; the check only distinguishes between ‘tool enabled’ and ‘not enabled’ and does not check whether updates are actually performed or merged (Fig. \ref{fig:single_checks}D).  \\
    % Code-review
    \textbf{Code-Review} identifies if a project enforces human code review before PRs are merged. 62.9\% of repositories received the lowest score of 0, while 3.8\% achieved the highest score of 10; the distribution shows that most projects have very low scores, with a low mean of 1.45 (Fig. \ref{fig:single_checks}E). Points were deducted if code changes were not reviewed by humans: 3 points deducted for unreviewed bot changes, 7 points for an unreviewed human change and a further 3 points if multiple human changes remained unreviewed; bot reviews did not count as valid code reviews.  \\
    %Maintained
    \textbf{Maintained} check assesses project activity through commit frequency and issue tracking. The distribution is bimodal with 63.5\% of projects scored 0 points as they were either archived or inactive, while 21.0\% scored full points for regular commits (at least one commit per week in the last 90 days) (Fig. \ref{fig:single_checks}F).   \\
    %Vulnerabilities
    \textbf{Vulnerabilities} check uses the OSV service (Open Source Vulnerabilities) \footnote{https://osv.dev/ (accessed 20/07/2025)} to identify unfixed known vulnerabilities. For each vulnerability found, the score was reduced by one. 62.8\% of the repositories achieved the highest score of 10, as they had no known, unfixed vulnerabilities in their own code or dependencies, while 21.8\% received the score 0 (Fig. \ref{fig:single_checks}G). \\  %0-9 mind. 1 vulerability \\
    %Binary Artifacts
    \textbf{Binary-Artifacts} check searches for binary artifacts in repositories that cannot be properly reviewed. With a high mean of 9.59, 91.8\% of the repositories achieved the score of 10, as they did not contain any binary files in the source code (Fig. \ref{fig:single_checks}H). \\
    %Dangerous-Workflow
    \textbf{Dangerous-Workflow} identifies three dangerous patterns in GitHub Action workflows, insecure use of pull\_request\_target or workflow\_run triggers and potential script injection attacks. 98.7\% of repositories did not contain such patterns (Fig. \ref{fig:single_checks}I).

\section{Discussion}
    %Summary (was ich gemacht habe  -> was bezwecken, was rausgekommen) 
    The current state of supply chain security in research software was investigated by analysing 3,248 GitHub repositories using the OpenSSF Scorecard. The results show an average score of 3.5 out of 10 (Fig. \ref{fig:general_scores}). 
    This is considerably below the OpenSSF-recommended threshold of 7 for adequate supply chain security \footnote{\url{https://best.openssf.org/SCM-BestPractices/github/repository/scorecard_score_too_low.html},(accessed 20/07/2025)}.
    For reference, similar aggregated scores have been observed in broader open-source ecosystems, with scores of 4.82 for projects in npm and 4.79 for PyPI derived from \citet{Zahan2023a} data. Though their security check scores were computed using an earlier Scorecard version and may not be directly comparable due to recent increases in scoring sensitivity for several checks.
    % Top Scores & Literature
    The top scoring high-risk checks are Binary-Artifacts, Vulnerabilities and Dangerous-Workflows (Figs. \ref{fig:single_checks}G-I). 
    This matches the results presented in \cite{Zahan2023a}, where the Binary-Artifacts and Dangerous Workflow checks also score consistently high in the scanned npm and PyPI ecosystems. The comparatively low Vulnerability score is a result of a software upgrade, that increased the sensitivity of this check. Older Scorecard versions only looked for directly reported vulnerabilities of the package whereas newer versions also check dependencies.
    % Lowest Scores & Literature
    Conversely, the lowest scoring high-risk checks are Signed-Releases, Token-Permissions and Branch-Protection (Figs. \ref{fig:single_checks}A-C). Here as well, one can find a similar scores for the Signed-Releases and Branch-Protection checks in \cite{Zahan2023a} although the score for Token-Permissions is notably worse. This difference can also be attributed to a change in the underlying scoring mechanism. 
    % Interpretation (Interpretation ergebnis, was bedeutet das?)
    The analysis reveals that most security-critical practices are either nor or only insufficiently implemented, evidenced by the fact that 6 out of 9 high-risk checks having a mean score below 3 (Fig. \ref{fig:single_checks}). 
    % Context
    The results indicate that research software repositories in their current form are inadequately protected against supply chain attacks. As compromised research software can result in data leakage or manipulation, it poses significant risks to the reproducibility, integrity, and reusability of scientific research, as well as to compliance with FAIR principles.
    The low adoption rate of security practices, including those with minimal implementation effort suggests a possible lack of security awareness, a possible lack of knowledge of source code management platforms' capabilities and a possible lack of resources to address these issues. However, these causes are currently speculative, targeted surveys among research software developers are needed to empirically identify the underlying causes.
    If confirmed, they could be addressed by increasing awareness and providing adequate training. As recommended by \citet{Anderson2020}, it is suggested that security aspects should be systematically integrated into software engineering and research software engineering courses.
    The OpenSSF Scorecard could be used as a practical teaching tool to illustrate security best practices, allowing students to directly assess and improve the security posture of their own projects.
    However, addressing these challenges should not rely solely on individual developers. Responsibility must also be shared with other stakeholders in the software ecosystem. For instance, platforms that publish research software could include selected security practices (e.g. branch protection, a security policy or signed releases) as part of the publication requirement. Git hosting platforms such as GitHub should also support security by making configurations secure by default \cite{CISA2023}. Furthermore, long-term funding from funding agencies is needed to ensure ongoing maintenance and sustainable security improvements for research software.
    % OpenSSF SCorecard
    The use of the OpenSSF Scorecard as supply chain security assessment tool enables an automated, comparable and reproducible evaluation of many projects \cite{OSSF_Scorecard}. However, there are methodological limitations. Some best practices such as two-factor authentication or multi-factor authentication are not machine-readable and therefore not captured. It also does not check quality of code review or the response times to vulnerabilities.
    % Outlook and future research
    Future studies should compare our findings with analyses based on alternative security tools and metrics to validate and extend these results.
    Additionally, future work should evaluate the feasibility and impact of the our prioritised recommendations through experiments, such as user studies with Research Software Engineers. The effort required, the relevant resources, the level of acceptance and the actual security gains should be measured. To better understand and categorise the specific weaknesses and strengths of research software in security, a systematic comparison with non-RS repositories is desirable. Such a comparison could also inform the development of a security maturity model specific to research software, marking an important next step.

\section{Recommendations}

    The OpenSSF Scorecard Security Checks were evaluated for risk and feasibility to provide actionable recommendations for RS engineers to improve supply chain security, using a risk-adoptability matrix (Fig. \ref{fig:check_matrix}) and implementation state of high impact security practices (Fig. \ref{fig:single_checks}). Based on the analysis, the following concrete measures are recommended for adoption by researchers, specifically targeting GitHub-centred projects:

    \begin{figure}
        \centering
        \captionsetup{font=footnotesize}
        \includegraphics[width=0.5\columnwidth]{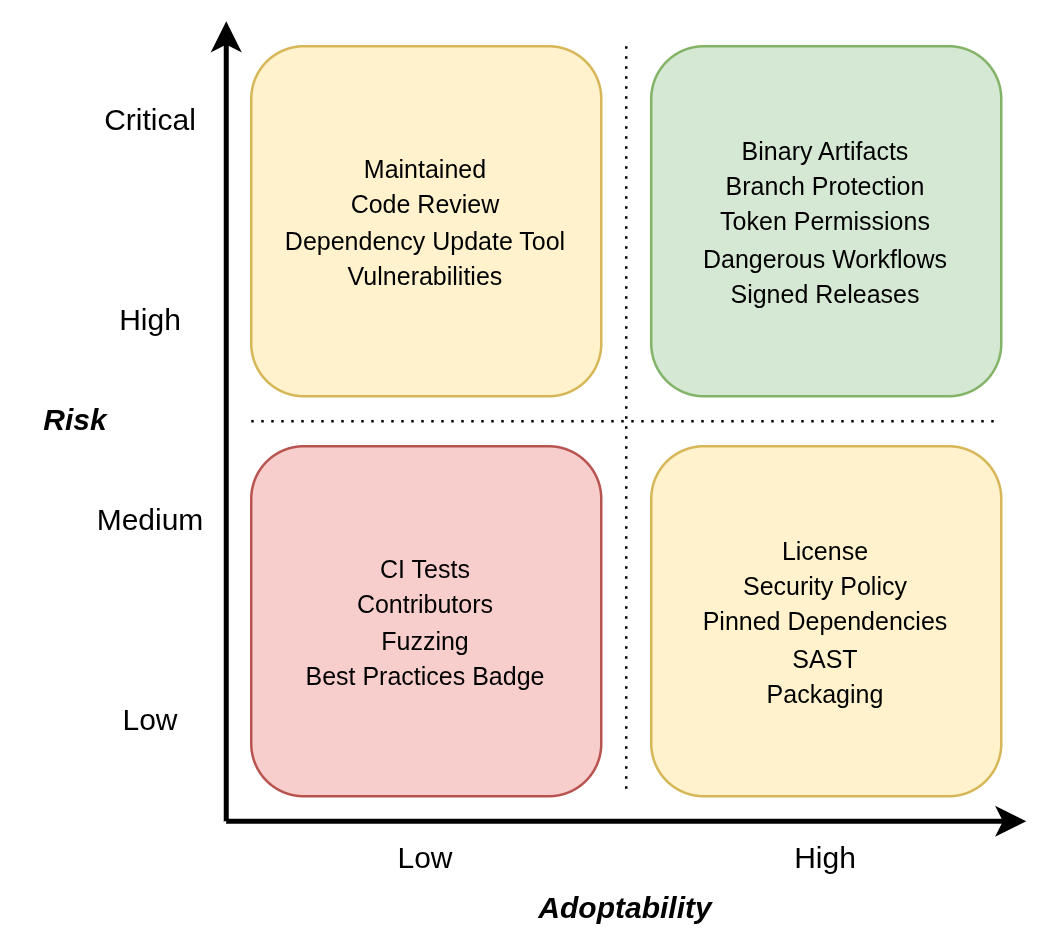}
        \caption{\textbf{Ranking Scorecard Checks.} Risk adjustment matrix for prioritising security measures for research software ranks the Scorecard checks according to the relationship between the security risk and the required implementation effort of the security practice for research software developers.}
        \label{fig:check_matrix}
    \end{figure}

    \textbf{Recommendation 1: Set up branch protection rules.} Modern SCM platforms such as GitHub or GitLab enable the maintainers of a repository to set up general rules for how contributors must interact with specific branches and tags. These rules will mostly follow standard patterns, with little to no maintenance overhead. Using this feature, it is recommended to block force pushes and the deletion of main development and release branches. For projects with multiple contributors, code reviews should be enforced by requiring a pull request for any changes on the main branch. \\
    \textbf{Recommendation 2: Use restrictive token permissions.} Many projects will opt to use inbuilt CI/CD features such as ''GitHub Actions'' for their projects, because they eliminate the need for open-source projects to set up and maintain their own build infrastructure. When using GitHub Action workflows, it is important to restrict the permissions of individual jobs as much as possible, to prevent attackers from circumventing security controls or exfiltrating environment variables and secrets from a repository. For many reusable workflows, the minimal permissions are stated in the documentation and just need to be copied. Additionally, automated tools exist to assist maintainers in securing their workflows \footnote{Step Security - Secure Workflow Tool: https://app.stepsecurity.io/secure-workflow (accessed 20/07/2025)}. \\
    \textbf{Recommendation 3: Sign your releases.} For many software packages, there is currently no reliable method for consumers to verify that a pre-built package has not been maliciously altered after uploading. To provide this assurance to users, maintainers should leverage the inbuilt ''Releases'' feature of GitHub and upload release assets for every published version of a package. The release assets should include the source code, a prebuilt package as well as signatures for every asset. Signatures can be created using established tools such as GPG. Recent developments, such as GitHub's artifact attestation feature, represent alternatives that are potentially easier to use for both maintainers and consumers \footnote{https://docs.github.com/en/actions/how-tos/security-for-github-actions/using-artifact-attestations/using-artifact-attestations-to-establish-provenance-for-builds (accessed 20/07/2025)}.

\section{Conclusion}
    In this study we performed an analysis of the supply chain security of quality research software. We find that most research software is inadequately protected against supply chain threats. This holds true for both high and low effort mitigations to known risks and thus points to a lack of knowledge of RS engineers on fundamental security practices as well as protection mechanisms of SCM platforms.

    \textit{The data and code supporting the findings of this article are openly available in Zenodo at https://doi.org/10.5281/zenodo.16279383}

%% Starten Sie "biber lni-paper-example-de", um eine Bibliographie zu erzeugen.
\printbibliography

\section*{Appendix}
        
% tabelle kann  eigentlich in den appendix da alle infos außer Std deviation in den plots sind
     \begin{table}[ht]
     %schriftgröse verkleinern
        \captionsetup{font=footnotesize}
        \centering
        \scriptsize
        %\rowcolors{2}{gray!15}{white} % Alternate row colors
        \begin{tabular}{lccl}
            \toprule
            %\rowcolor{white} % Ensure header row is white
            \textbf{Security Check} &  \textbf{Mean Score} & \textbf{Std Dev} & \textbf{Risk Level} \\
            \midrule
            Fuzzing  & 0.04 & 0.61 & Medium \\
            CII-Best-Practices & 0.08 & 0.50 & Low \\
            Pinned-Dependencies & 0.11 & 0.67 & Medium \\
            Signed-Releases  & 0.20 & 1.26 & High \\
            Security-Policy  & 0.27 & 1.58 & Medium \\
            Token-Permissions  & 0.30 & 1.70 & High \\
            SAST  & 0.35 & 1.72 & Medium \\
            Branch-Protection  & 1.39 & 2.39 & High \\
            Code-Review  & 1.45 & 2.65 & High \\
            Dependency-Update-Tool  & 1.45 & 3.52 & High \\
            Packaging  & 1.46 & 3.53 & Medium \\
            Maintained  & 2.71 & 4.12 & High \\
            CI-Tests & 3.79 & 4.41 & Low \\
            Contributors  & 6.20 & 3.97 & Low \\
            Vulnerabilities & 7.07 & 4.23 & High \\
            Binary-Artifacts & 9.59 & 1.73 & High \\
            Licence & 9.65 & 1.43 & Low \\
            Dangerous-Workflow & 9.87 & 1.13 & Critical \\
            \bottomrule
        \end{tabular}
        \caption{\textbf{Scores of Single Security Checks of Repositories}. For each security check, the mean and standard deviation of repository scores (excluding -1) are shown, along with the corresponding risk level as classified by the OpenSSF Scorecard.}
        \label{tab:security-checks}
    \end{table}

\end{document}